\documentclass[preprint,12pt]{elsarticle}
\usepackage{tabularx}

\usepackage{titlesec}
\usepackage{comment}
\titleformat{\paragraph}
  {\normalfont\normalsize\bfseries}{\theparagraph}{1em}{} 
\titlespacing*{\paragraph}{0pt}{3ex plus 1ex minus .2ex}{1ex}

\usepackage{amssymb}
\usepackage{amsmath}

\usepackage{hyperref}

\begin{document}

\begin{frontmatter}



\title{Managing Security Issues in Software Containers: From Practitioners' Perspective}





\author[label1]{Maha Sroor}
\author[label1]{Rahul Mohanani} 
\author[label2]{Ricardo Colomo-Palacios} 
\author[label3]{Sandun Dasanayake}
\author[label1]{Tommi Mikkonen}


\affiliation[label1]{organization={University of Jyväskylä, Faculty of Information Technology},
            addressline={Mattilanniemi 2}, 
            city={Jyväskylä},
            postcode={40100}, 
            state={},
            country={Finland}}
\affiliation[label2]{organization={Technical University of Madrid},
            addressline={Calle Los ciruelos}, 
            city={Boadilla del Monte},
            postcode={28660}, 
            state={Madrid},
            country={Spain}}   

\affiliation[label3]{organization={University of Oulu},
            addressline={Pentti Kaiterankatu 1}, 
            city={Oulu},
            postcode={90570}, 
            state={},
            country={Finland}}            

\begin{abstract}

Software development industries are increasingly adopting containers to enhance the scalability and flexibility of applications. Security in containerized projects is a critical challenge that can lead to data breaches and performance degradation, thereby directly affecting the reliability and operations of the container services. Despite the ongoing effort to manage the security issues in containerized projects in SE research, more investigations are needed to explore the human perspective of security management to security management in containerized projects. This research aims to explore security management in containerized projects by exploring how SE practitioners manage the security issues in containerized projects. A clear understanding of security management in containerized projects will enable industries to develop robust security strategies that enhance software reliability and trust. To achieve this, we conducted two semi-structured interview studies to examine how practitioners approach security management. The first study focused on practitioners' perceptions of security challenges in containerized environments, where we interviewed 15 participants between December 2022 and October 2023. The second study explored how to enhance container security, with 20 participants interviewed between October 2024 and December 2024. Data analysis reveals how SE practitioners address the various security challenges in containerized projects. Our analysis also identified the technical and non-technical enablers that can be utilized to enhance security in containerized projects. Overall, we propose a conceptual model that visualizes how practitioners manage security issues in containerized projects. We argue that our proposed model will guide practitioners in making informed decisions to plan, develop, and deploy secure container systems.

\end{abstract}




\end{frontmatter}



\section{Introduction}
\label{Introduction}


In recent years, the reliance on digital services has significantly increased across multiple sectors, driving significant advancements in digital transformation. Software-intensive industries such as finance, healthcare, manufacturing, retail, and government services have increasingly integrated digital technologies to enhance efficiency, security, and scalability. \cite{liva2020exploring} \cite{maltese2018digital}.
Similarly, many public and private organizations are increasingly adopting software applications to optimize operations and enhance customer experience  \cite{almeida2020challenges}. This growing dependence on software applications and the incremental customer demand for new features has further added to the complexity of software applications, introducing new challenges in designing, developing, testing, and deploying safe and reliable software applications. To overcome this issue, containerization has emerged as a credible solution by improving flexibility, portability, and agility in software development and deployment. Containerization encapsulates applications and their dependencies into isolated environments, ensuring consistent service delivery across different infrastructures \cite{koskinen2019containers}.

Despite many advantages over Virtual Machines (VMs), \cite{benguria2018agile}, software containers (mentioned as only `containers' henceforth) introduce significant security concerns, such as Faulty image, vulnerable configurations, unauthorized access, and data leakage \cite{vs2023container}. These security concerns are considered the main barriers to a wider container adoption \cite{combe2016docker}. A thorough investigation of security issues in containers and their implications is critical for organizations because it would help them improve their security strategies and ensure the continuous delivery of software services \cite{sultan2019container}. Furthermore, investigating security practices in containerized environments can enhance trust and encourage broader adoption \cite{sultan2019container}.



A review of existing software engineering (SE) literature on security concerns in container systems reveals a lack of a holistic view of security management in containerized projects. Many studies discuss security challenges in containers \cite{martin2018docker} \cite{kaur2021analysis} but do not detail the practical challenges that obstruct security in containerized projects. Although some studies propose frameworks to manage and improve security in container systems \cite{mahavaishnavi2024secure} \cite{chen2022seaf} \cite{jolak2022conserve}, the frameworks often remain impractical for real-world projects as they do not align with domain-specific requirements. Moreover, studies do not consider the non-technical factors that affect container deployments, such as human administration, communication, budget constraints, and customer demand \cite{sroor2024practitioners}. These security frameworks do not incorporate the human perspective in managing container systems, neglecting the significant role of humans in container administration.

Software containers are not fully automated, and they need humans to administrate the technical aspects, such as configurations, monitoring, and version control. Usually, security management in container systems focuses on configuration and testing practices and often does not consider the human perspective \cite{siddiqui2019comprehensive} \cite{souppaya2017application} \cite{balzacq2010security} \cite{shamim2020xi}. Including the human aspect in managing security is crucial because it adds new dimensions to security management, such as planning, strategic decision-making, policy enforcement, and continuous adaptation to new threats, making security management in containers more beneficial in practice. A comprehensive understanding of security management in containerized projects will enable industries to develop strategies that enhance software reliability.

Consequently, this research aims to provide an in-depth analysis of practitioners' understanding of security management in containerized projects. It extends our previous work ( as reported in \cite{sroor2024practitioners}) that analyzes risks and vulnerabilities in software containers in addition to their causes and implications from the point of view of how SE practitioners understand security issues in containerized projects. Furthermore, our research highlights container security's weaknesses and strengths from the practitioners' perspectives. The current study extends our previous findings by understanding how SE practitioners manage these critical issues and concerns in containerized projects. Hence, the main research question guiding this paper is: 

\noindent \textbf{\textit{How do SE practitioners manage security challenges in containerized projects?}} 

To answer our main research question, two main aspects need to be investigated. The first aspect is to explore how practitioners perceive security issues in containerized systems in terms of their causes and implications. The second aspect is investigating how SE practitioners manage these security issues in containerized projects.

This research makes the following novel contributions to the scientific body of knowledge on managing container security in containerized projects:

\begin{enumerate}
    \item Provides a comprehensive analysis of the perception of container security from the perspective of software practitioners; 
    \item Identifies the key strengths and weaknesses of container security practices, highlighting real-world challenges encountered in containerized projects;
    \item Explores the fundamental (or key) enablers that can enhance security in containerized environments by combining the technical and non-technical factors influencing container security management.

    \item Contributes a conceptual model to guide SE practitioners toward developing robust strategies for managing security in containerized projects.
\end{enumerate}

The paper is structured as follows. Section \ref{Background} provides a comprehensive overview of software containers and container security. Section \ref{Study Design} describes the overall research process employed to collect and analyze the data for this study. Section \ref{Results} reports the main findings of the study. Section \ref{Discussion} interprets the findings and their reflection on practitioners and future research. Section \ref{Conclusion} concludes the study by highlighting the main contributions.

\section{Background}
\label{Background}
This section introduces the background of this research, giving an overview of software containers, security issues in container systems, and managing security in container systems. 

\subsection{Software Containers }

Containers are lightweight, executable software packages that encapsulate an application along with its dependencies, including libraries, configuration files, and binaries, ensuring consistent execution across different environments \cite{7093032}. Unlike virtual machines (VMs) that run their operating system on top of the host system, containers share the host operating system kernel to improve resource utilization \cite{felter2015updated}. Additionally, containers operate in isolated environments to enhance the security and stability of software applications \cite{bentaleb2022containerization}.

The development of software containers begins with creating a container image. Container images are built in-house or pulled from private registries managed by organizations or public registries\cite{kaur2021analysis} such as Docker Hub \footnote{https://hub.docker.com/} or Amazon ECR Public Gallery \footnote{https://gallery.ecr.aws}. After pulling the image, the hosting machine is configured to allocate system resources, including memory, CPU(s), and file access, to ensure optimal performance \cite{bernstein2014containers}. Additionally, network segmentation and configuration are required to enable secure communication between containers and external systems \cite{hoenisch2015four}.

 Deploying the software containers requires a hosting machine to execute the application in the container image. The container creates an instance of its image and operates in an isolated environment to \cite{hoenisch2015four}. Container orchestration tools like Kubernetes \footnote{https://kubernetes.io/} manage the deployment, scaling, and operation of containers in the cluster to provide the service \cite{rahman2023security}.

Deploying software containers involves executing the application encapsulated within the container image on a hosting machine. The container instantiates an image and operates within an isolated environment to prevent conflicts with other applications \cite{hoenisch2015four}. Then, the orchestration tools such as  Kubernetes \footnote{https://kubernetes.io/} automate deployment, scaling, and operational management, ensuring high availability and efficient resource allocation within a cluster \cite{rahman2023security}.

Containers offer numerous advantages to software deployment \cite{sroor2022leverage}. One of the main advantages is their ability to provide portable environments across different stages of development, testing, and production \cite{benguria2018agile}. Containers are lightweight and consume fewer resources compared to VMs. Containers are also scalable; they can be scaled up or down according to resource demand \cite{benedicic2019sarus}. Additionally, containers support microservices architecture, allowing developers to break down applications into smaller, manageable components \cite{reifer2002good}.

There are many containerization technologies, such as LCX, Docker, Podman and others. LCX is one of the early container technologies that are used run to run multiple isolated Linux systems \cite{bentaleb2022containerization}. Docker is one of the most popular container technologies. It provides a simplified creation, deployment, and running of containers \cite{kithulwatta2021adoption}. Podman is also a leading container technology. It is used in high-performance computing (HPC) environments as it offers a daemon-less container engine \cite{gantikow2020rootless}.


\subsection{ Security Issues in Software Containers}

Security management in containers is considered one of the biggest challenges for this technology \cite{vs2023container}. Considering container security challenges is crucial to improving container adoption and encouraging container migration \cite{sultan2019container} \cite{martin2018docker}. Consequently, there is a need to comprehend the container security issues and management as they affect usability, performance and service availability \cite{kaur2021analysis}.

 Delivering secure service by containerized applications requires embedding security as an element in the development and deployment of containers. This happens in multiple phases. The first phase is the container image, a lightweight and executable software package essential for running containers \cite{xu2018mining}. The second phase is the container host preparation, including the infrastructure settings to provide the container with an isolated environment \cite{bernstein2014containers}. The third phase is intra-containers, which is a running form of the image and lightweight virtualization of the software \cite{paraiso2016model}. The fourth phase is networking, which facilitates communication with external entities and establishes internal communication channels \cite{casalicchio2019container}. The fifth and last phase is Runtime, which is a tool used to manage and execute containers to deliver the service in production  \cite{ibrahim2019attack}.  The security decisions within these phases are critical, as they frame the techniques by which security should be implemented and configured. 


Security issues often arise due to configuration flaws and inadequate security practices during container deployment \cite{zerouali2019impact}. In the container image phase, security threats primarily stem from malicious image sources and insufficient vulnerability scanning. Pulling container images from untrusted registries allows malicious codes to cause harm to the container system. Ignoring or insufficient image vulnerability scanning exposes sensitive credentials \cite{shu2017study}.

Security issues in the container host affect isolation and data protection in container systems \cite{martin2018docker}. It basically happens because of insecure configurations, inefficient resource isolation, and host-escalated access permissions. An insecure container host exposes the container system to buffer overflow, host exhaustion, unauthorized access, and data breaches \cite{gantikow2016providing} \cite{dissanayaka2020vulnerability}.


The main causes of security issues within containers are unauthorized access, misconfiguration, and weak isolation mechanisms\cite{sultan2019container}. These causes may lead to Denial-of-Service (DoS) attacks or complete system failures \cite{jian2017defense} \cite{mp2016enhancing}. Some containerized applications rely on dynamic architectures to manage workload scaling. If scaling is not properly managed, it may lead to host exhaustion \cite{xu2019dockerfile}.

Security issues in container networks and orchestration usually occur because of insecure external communication with other systems or insecure internal communication between the clusters. Misconfiguring the network or the orchestration settings leads to network and orchestration security issues. Misconfigurations can lead to a privileged network, allowing unrestricted capabilities to the network. Consequently, unauthorized access to the network means controlling the nodes (servers, routers, or devices) \cite{budigiri2021network}. Another reason is the poor segregation of the container network, which might expose sensitive data or make it vulnerable to intercepted network traffic \cite{budigiri2021network}.

Security issues during runtime occur because of the development and production implementation in the same physical environment. It is risky behaviour as it increases surface attacks \cite{martin2018docker}. Runtime misconfiguration also exposes the container’s potential performance issues that affect stability and service delivery \cite{gholami2021should}.

\subsection{Managing Security in Container Systems}

security management in container systems is crucial to protect container applications. One of the security management approaches is testing. Container testing is essential for ensuring the security, efficiency, and reliability of containerized applications. Container testing is the process of detecting anomalies that could disrupt the progress of containerized software development. Container testing encompasses analyzing container images, configurations, container communication, and pipelines \cite{siddiqui2019comprehensive}. Implementing comprehensive testing protocols is essential to ensure the robustness and continuation of container functionality as well as maintain the integrity and reliability of applications deployed in containers \cite{souppaya2017application}.

Another approach to managing container security is implementing security practices. Security Practices refer to enhancing the security of container systems through collective processes and techniques \cite{balzacq2010security}. One of the primary best practices is to use trusted base images and regularly scan them for vulnerabilities \cite{haque2022well} \cite{doan2022davs}. This helps in minimizing the risk of introducing security flaws into the container environment. Additionally, employing role-based access control (RBAC) ensures that only authorized users can have access \cite{shamim2020xi}. Network segmentation and restricted network capabilities are crucial practices to limit communication between containers and avoid surface attacks \cite{budigiri2021network}.

Maintaining container systems involves continuous monitoring of container behaviour for anomalies that could indicate a security breach. Regular checking of logs can protect container runtime against known vulnerabilities \cite{candido2021log}. Effective secrets management is also critical to secure sensitive information. Secrets should not be accessible to all containers, and it is recommended to be stored in external volumes \cite{mondal2022kubernetes} \cite{sroor2024practitioners}. Regular security audits and compliance checks help ensure container development and deployment follow the internal and legal policies to ensure users' data privacy \cite{shamim2020xi} \cite{belair2021snappy} \cite{fernandez2019secure}.




Security in container systems faces significant issues in the development and deployment life-cycle. These challenges can potentially compromise the integrity and functionality of containerized applications. Security issues arise from faulty images, misconfigurations in the host machine, network settings, or container pipelines. Additionally, unauthorized access during runtime can further increase security risks, potentially leading to breaches or service disruptions. Therefore,  effective security management is essential to maintain the reliability, availability, and integrity of containerized applications and their services. A review of SE literature on security management in container systems shows that container security primarily relies on security practices and rigorous testing. However, human administration can significantly influence container security, there is a lack of consideration of the human role in planning, decision-making, and strategy development of security management in container systems. Thus, there is a critical need to explore the human perspective in security management to enhance the effectiveness and adoption of security practices within containerized environments.

\section{Study Design}
\label{Study Design}

This section outlines the study methodology, detailing the planning, data collection, data transcription process, and data analysis approach.

\subsection{Research Questions}
\label{Research method}

Improving security in containerized environments requires a thorough understanding of how security issues are conceptualized, managed, and implemented in software container projects. Exploring container security management practices will also help to strengthen security strategies and inform decision-making in container security. Accordingly, this study addresses the following main research question:

\noindent \textit{\textbf{How do SE practitioners manage security challenges in containerized projects?}} 

To comprehensively understand and explore how practitioners approach security management in containerized projects, it is necessary to explore how practitioners perceive the security issues, causes and implications in container systems. It also requires an examination of the different ways in which security can be improved in containerized projects. Therefore, 
the main research question is divided into two sub-research questions. The sub-research questions are as follows:

\noindent \textit{\textbf{RQ1}: How do SE practitioners perceive security issues in software containers?}

\noindent \textit{\textbf{RQ2:} How do SE practitioners address security issues in containerized projects?}

To address these research questions, we conducted two separate interview-based studies---Study 1 and Study 2---addressing RQ1 and RQ2, respectively. Below, we describe the research method employed for each study in more detail.

\subsection{Study 1}

\subsubsection{Research Approach}
\label{Research Approach}

We conducted a qualitative interview-based study (as suggested in \cite{hove2005experiences}) to answer RQ1---\textit{``How do SE practitioners perceive security issues in software containers?''}. A semi-structured interview guide was developed, following the established guidelines suggested by \cite{dicicco2006qualitative}. These guidelines provided a framework for ensuring flexibility and consistency across interviews. To maintain the reliability of the research method and ensure the quality of the interview protocol, we followed the guidelines for conducting interviews in \cite{strandberg2019ethical} and empirical standards for interview studies by ACM SIGSOFT \cite{ralph2020acm}. Furthermore, we followed the guidelines in \cite{saldana2021coding} for coding qualitative data and followed the process suggested in \cite{cruzes2011recommended} for thematically analyzing the findings. 

\subsubsection{Study Planning}

We recruited participants with at least one year of experience working with software containers in development, deployment, or managing containerized projects. The participants were recruited through a consortium for a project on containers \textit{Quantum Leap in Software Development (QLeap)}. We also contacted software practitioners from the research team connections using LinkedIn \footnote{https://www.linkedin.com}.

The interviews were conducted in English between December 2022 and October 2023 using the Microsoft Teams platform\footnote{https://www.microsoft.com/en-us/microsoft-teams}. Before beginning the interviews, participants were reminded of the main objective of the research. Participants agreed to start the recording, and the interviews lasted an average of 60 minutes. 

The interview guide consisted of two sections. The questions in the first section collected participants’ demographic data, such as the participant’s current country of employment, job title and related responsibilities, years of experience working with software containers, and the domains in which participants developed container applications. The second section comprised open-ended questions to explore container security issues, their causes, and relevant solutions in containerized projects. The interview guide is available here: \url{https://zenodo.org/records/10949260}.

\subsubsection{Piloting the Interview Instrument}

Two pilot interviews were conducted to improve the interview questionnaire. The primary objective was to evaluate the clarity and relevance of the questions and determine if participants could understand the core of the questions. Feedback from the participants was positive, thus validating the clarity of the questions. The pilot data was also analyzed to assess the quality and reliability of the data for final analysis. However, the data from the pilot interviews were not included in the final data analysis.

\subsubsection{Interview Sampling}

We recruited a total of fifteen participants using convenience sampling \cite{baltes2022sampling}, ensuring a diverse range of expertise in containerized software development and security. The participants held various roles in the software industry, including CEOs, security specialists, and software engineers, providing insights from both technical and strategic perspectives. To ensure diversity in organizational culture, security practices, and regulatory environments, we recruited participants from Finland, India, Sri Lanka, and the Netherlands.

Participants had IT industry experience ranging from three to twenty-eight years, with an average of eleven years, representing varying levels of seniority and expertise. Their experience working with containerized applications ranged from one to eight years, averaging approximately four years, ensuring a well-rounded perspective on container security challenges and best practices. The demographic diversity of the sample (as summarized in Table \ref{Participants Demographic Data}) was intended to enhance the generalizability of the findings across different roles, industries, and organizational contexts within the software sector.

\begin{table}[htp]
\centering
\scriptsize
\begin{tabular}
{p{1cm}p{2cm}p{4cm}p{2cm}p{3cm}}  
\textbf{ID} & \textbf{Country} & \textbf{Role}&\textbf{Experience}& \textbf{Domain} \\ 
P1 & Finland & Developer & 5 & Higher Education \\ 
P2 &Finland  & Senior SW Engineer & 6 & Gaming \\
P3 &Finland  &  CTO & 5 & Web Applications \\
P4 &Finland & Security Delivery Specialist & 2 & IoT \\
P5 &Finland & lead Architect & 5 &  Healthcare \\
P6 &Finland  & Security Engineer & 3 & Elevators \\
P7 &Finland  & Team supervisor & 6 & Telecommunications \\
P8 &Finland & Senior SW Engineer & 4 & E-commerce \\
P9 &Sri Lanka &  DevOps Engineer & 1 & Fintech \\
P10 &Sri Lanka & CEO & 6 & logistics \\
P11 &India & CTO & 8 & Electronic Medical Record \\
P12 &India & Cloud Architect & 7 & Telecommunications \\
P13 &Finland & DevOps Engineer & 3 & Web Applications \\
P14 &Netherlands & Testing Engineer & 4 & Healthcare \\
P15 &Finland & Senior SW Architect & 8 & Healthcare \\

\end{tabular}
\caption{Participants' demographic data Study 1} 
\label{Participants Demographic Data}
\end{table}

\subsubsection{Data Transcription and Management}

The audio files were transcribed into text using the automated feature in Microsoft Teams. The first author checked all 15 transcripts manually to ensure the transcribed data reflected the audio. Then, the entire author team validated the accuracy of the transcription process by randomly choosing 3-5 transcripts. The transcribed interview files were completely anonymized and renamed into identifiers numbered from P1 to P15 to ensure transcripts could not be traced back to reveal the participants' identities. The files were uploaded to ``Atlas.ti' \footnote{https://atlasti.com}, which is known for its advanced coding capabilities that facilitate organizing, analyzing, and visualizing qualitative data.

\subsection{Study 2}
\subsubsection{Research Approach}

We conducted a qualitative interview-based study (as suggested in \cite{hove2005experiences})  to answer RQ2---\textit{``How do SE practitioners manage security issues in containerized projects?''}. We followed the same approach and guidelines for developing interviews as employed in \textit{Study 1} (refer to \ref{Research Approach}).


\subsubsection{Study Planning}


Participants were recruited through LinkedIn connections, research groups from other partnering universities, and industries collaborating with the research team. We targeted participants with substantial experience in developing and deploying software containers. The interviews were conducted in English between October 2024 and December 2024 using the Microsoft Teams platform. Before beginning the interviews, participants were reminded of the study's main objective. The interviews lasted for an average of 48 minutes.

Demographic data was collected using an online survey to optimize the interview duration, which included questions about participants' country of employment, job title, number of containerized projects they had participated in, and the work domain. The survey is available here: \url{https://link.webropolsurveys.com/S/EFC982BAD0C6E07E}. 
The interview guide had open-ended questions that collected data about security practices, testing, logging and monitoring, and human communication. The interview guide was shared with interviewees one day before the interviews. This gave them sufficient time to review the questions and think about responses to maintain informed discussion. The interview guide is available here: \url{https://zenodo.org/records/14645107}.

\subsubsection{Piloting the Interview Instrument}

We conducted two pilot interviews to fine-tune and optimize the interview guide. The participants were recruited from industries collaborating with our research team, and the interviews were recorded and analyzed to assess the data quality. 

Based on feedback from pilot participants, two interview questions were reworded for improved clarity and precision. The first question---\textit{How can logging and monitoring help container security?} was revised to---\textit{In your opinion, how can logging and monitoring help manage container system security?} to encourage a more nuanced response regarding security management practices. Whereas the second question---\textit{How can AI play a role in security practices to support container security?} was refined to---\textit{How can AI be embedded in current practices to improve container security?} to emphasize the integration of AI within existing security workflows.


\subsubsection{Interview Sampling}
We sampled a total of 20 interviewees. Participants have various roles, including project coordinator, software designer, software engineer, developer, tech lead, developer, architecture engineer, team manager,  researcher, post-doctoral researcher, and university professor to include various experiences and backgrounds in software container development and deployment. Participants were recruited from Finland, Spain, Sri Lanka, India, Colombia, Poland, and the Czech Republic to ensure organizational culture and diversity standards. The participant's experience in containers ranged from 10 to 1 year, and the number of projects ranged from 20 to 1 project, with an average of 3.5 years of experience and an average of 4 projects per participant to ensure various levels of experience. The diversity of the participants (as shown in  Table \ref{Participants' demographic data Set2})  was aimed to ensure the generalizability of the findings across various roles, experiences and domains within the software industry.

\begin{table}[htp]
\centering
\scriptsize
\begin{tabular}{p{1cm}p{2cm}p{3.5cm}p{2cm}p{3.5cm}}
\textbf{ID} & \textbf{Country} & \textbf{Role} & \textbf{Experience} &\textbf{Domain }\\ 
I1 & Finland & Researcher- Developer & 2 & Edge computing \\ 
I2 &Finland  & Project coordinator & 2 & Education\\
I3 &Finland  &  Researcher & 3 & Software ecosystems \\
I4 &Finland & Doctoral researcher- Architecture Engineer & 3 & Cloud computing platform \\
I5 &Finland & Project researcher & 2 & Academia \\
I6 &Finland  & Software designer & 1 & Web service \\
I7 &Spain & Software Engineer & 2 & Order management \\
I8 &Poland & Doctoral researcher- Team manager & 5 & Web service \\
I9 &Spain & Backend Software Engineer & 1 & Healthcare \\
I10 &Portugal & Application Security Consultant & 4 & Telecommunications \\
I11 &Estonia & Developer & 3 & Logistics \\
I12 &Czech Republic & Web Developer & 2 & Web Development \\
I13 &Colombia & IT Project Manager - Professor & 10 & E-commerce \\
I14 &Finland & Postdoctoral Researcher & 8 & E-commerce \\
I15 &Spain & Associate Professor & 5 & Digital literacy \\
I16 &Spain & Researcher- Developer & 2 & Bioinformatics \\
I17 &India & Software Engineer & 3 & Web service \\
I18 &Sri Lanka & Software Engineer & 2 & Manufacturing \\
I19 &Sri Lanka & Tech Lead & 3 & Web service\\
I20 &Sri Lanka & Software Engineer & 1 & Manufacturing\\
\end{tabular}
\caption{Participants' Demographic Data Study 2} 
\label{Participants' demographic data Set2}
\end{table}

\subsubsection{Data Transcription and Management}
The interview files were transcribed into text using the automated feature in Microsoft Teams. The first author checked all 20 transcripts manually to ensure the transcribed data reflected the audio. The author team validated the accuracy of the automatic transcription process by randomly choosing 3-4 transcripts. The transcribed interview files were completely anonymized and renamed into identifiers numbered from I1 to I20. After renaming, there was no way in which the transcripts could be traced back to reveal the identities of the participants. The files were uploaded to ``Atlas. ti' to facilitate further analysis.

\section{Data Analysis }
This section presents qualitative data analysis. We employed the thematic analysis approach suggested by \cite{cruzes2011recommended},  which allows for the systematic identification, analysis, and reporting of patterns (themes) within qualitative data.

\subsection{Study 1}
\label{study 1}

\subsubsection{Familiarization with the data}
The first author carefully read all 15 interview transcripts, ensuring familiarity with the participants' responses and gaining a comprehensive understanding of the content. This process facilitated the initial recognition of the main ideas and potential themes from the data. The research team discussed potential coding schemes to maintain rigour and consistency in the coding approach. The research team confirmed that all relevant aspects of container security were adequately captured.

\subsubsection{Generating Codes}

Once the data was familiarized, the first author systematically coded all the transcripts using ``Atlas.ti'' line-by-line to ensure that each text segment was analyzed in detail. The coding process resulted in 227 data segments, and descriptive codes were assigned to each segment. To enhance the reliability of the coding process, the second author independently reviewed and validated the assigned codes, ensuring that they accurately reflected the content of the transcripts. The author team discussed and approved the coding process. 

\subsubsection{Forming Themes}
To increase the level of abstraction, the identified codes were grouped into themes. The themes are a high-level conceptualization of multiple codes grouped together to describe a significant aspect of practitioners' experiences with container security. The first author conducted the thematic analysis. The themes that emerged were as follows---\textit{Experience-based knowledge, Container Security as a Chain of Dependencies, Preferring Automation, A Common Understanding of the Security Issues, Non-Technical Causes, Reliance on tools, Uncertainty about Improving Security Practices, Lack of Standardization and Guidelines, Unclear Resilience Time, and Container Security is Conditional}. The second author then audited the themes. The auditing resulted in renaming some of the themes (e.g., \textit{not preferring manual process} became \textit{preferring automation}). Detailed information about the codes and themes is in section \ref{Practitioners perspective on container security}. 

\subsubsection{Developing the Model}

Building on the identified themes, we developed an initial conceptual model (as shown in Fig. \ref{fig:model})  describing the interconnections between the key themes of the practitioners' perspective on container security concerns. This model was developed to illustrate how different aspects of container security interact and influence each other. After multiple discussions and refinements, the author team finalized the model.

\subsection{Study 2}

\subsubsection{Familiarization with the data}
 The first author read 20 interview scripts while ensuring the quality of the manual transcription. Due to the large volume of data, the first and second authors conducted a second round of reading to deepen their understanding of the data and refine the initial coding ideas. The research team discussed potential coding schemes to maintain rigour and consistency in the coding approach. The author team discussed and confirmed the coding ideas. 

\subsubsection{Generating Codes}
The coding process was conducted systematically to ensure a rigorous and reproducible approach to data analysis. The first author employed Atlas.ti, to facilitate the coding process.  The coding process resulted in 310 data segments, and descriptive codes were assigned to each segment. We realized that our analysis reached data saturation after the sixteenth interview when no new codes emerged. To ensure the reliability and validity of the coding scheme, the second author conducted an independent validation by reviewing a subset of the coded data. The entire author team discussed and approved the coding process.

\subsubsection{Grouping Codes into Themes}
The first author grouped the relative codes into categories to describe how practitioners managed security concerns in software containers. The following themes emerged---\textit{ Improving the Knowledge about Container Security, Human Collaboration and Communications, Artificial Intelligence, Security Practices, Risk Identification, Container Testing, and Logging and Monitoring}. The second author reviewed the themes and suggested a higher level of abstraction for the identified themes. We further identified the emerged themes as fundamental (or key) enablers of container security from the point of view of SE practitioners.

The themes, or the key enablers, were further categorized into \textit{technical enablers} of container security and \textit{non-technical enablers} of container security. The themes and the higher level of abstraction for the themes were reviewed and confirmed by the entire author team. Detailed information about the codes, themes, and further categories are provided in \ref{Key enablers for container security} section. 

\subsubsection{Developing the Model}

The final step of the thematic analysis produced our revised model by combining the findings from the first study, as shown in Fig. \ref{Key Improvements in Container Security}. This model presents a visual presentation of how practitioners manage security in containerized projects. Model 2 integrates the themes identified in \textit{Study 1 }(see Section \ref{study 1}) with those from \textit{Study 2} to a comprehensive view of the interrelationships between the key aspects of container security and the enablers for security improvement.

\section{Results}
\label{Results}

In this section, we report the findings from analyzing the data collected from interviewing the practitioners regarding managing container security in containerized projects. 

\subsection{Study 1: Practitioners' Perspective on Container Security in Practice}
\label{Practitioners perspective on container security}

To manage security issues in containerized projects, we must first understand and explore how practitioners perceive security issues in containers. This subsection focuses on the security patterns of container security issues, security issues, and implications from practitioners in containerized projects. The identified patterns are ordered from foundational concepts, such as practitioner knowledge and employed practices, to more advanced themes, such as standardization and practitioners' opinions about container security. All the quotes' themes, codes, and samples are summarized in table \ref{A summary of the Thematic Analysis Study 1}. A consolidated spreadsheet with data analysis, including all the themes, codes, and quotes, is available at: \url{https://zenodo.org/records/10959273}. More detailed descriptions of each theme are provided across the next few sections.

\begin{table}[htp]
\centering
\scriptsize
\begin{tabular} {p{3cm}p{3cm}p{9cm}}
\textbf{Theme} &\textbf{Example Codes} & \textbf{Example  Quotes} \\
Experience-based  knowledge &Classic issues & “Excluding  container  escapes;  everything  is  just  a  reiteration  of classical problems.” (P6) \\
 &Well-known issues & “Normally, we don’t have any new security issues. ”(P15) \\
 &Unknown issues & “We didn’t know this container problem or where it comes from, we
killed it, hoping everything will go safely.”(P13) \\
Container security is a chain of dependencies  &Containers are integrated
pieces & “If one part of the container goes down, it might cause problems with
the others as well.” (P14) \\
 &Delayed security implications  & “  Whenever we pull images, we don’t know if it is actually secure,
and we will never know until something bad happens.”(P2) \\
Preferring automation &Automation preference & “  If there are any security automated tools, they will be better than
humans.” (P13) \\
 &Human mistakes & “Usually developers want to do tasks as fast and easy as possible,
meaning insecure shortcuts in most cases.” (P3) \\
A common understanding of the security issues &Image issues & “Yeah, if you’re using the outdated base image, there will be vulnerabilities that need to be fixed.”  (P1) \\
 &Host issues & “  Something that  you need  to  be  aware  of is  memory and  CPU reservation to avoid container exhaustion.” P2)  \\
 &Intra-Container issues & “Usually, developers don’t settle container privileges; they concentrate
too much on security issues.”   (P12 ) \\
 &Network issues & “The first issue that I face is container ingress and egress ports. Ports
are left open for the entire world ”   (P12) \\
 &Recurring security issues & “Ohh, the most recurring  issues, misconfigured containers.”  (P6 ) \\
Non-technical causes &Technical causes & “I  think main causes are lack of knowledge and tooling to scan
containers, applications, and codes.” (P4 ) \\
 &Managerial causes& “It takes time to train a new developer to your work processes.”  (P2) \\
Reliance on tools &Configuration management  & “We  are  planning  to  use  Red Hat  Advanced Cluster  Security for
Kubernetes .”   (P10) \\
 &Code quality & “ We use Coverity to identify code quality issues.”   (P6 ) \\
 &Monitoring & “Grafana  Dashboard and Prometheus are used for monitoring system
metrics.” (P9) \\
Uncertainty about improving security practices &Volumes conflict & “Keep the container stateless and use volumes to reduce complexity.” (P12 ) VS “If you have sensitive data, you can’t expose it to access to the container.”  (P2)  \\
 &Layered approach conflict & “I  would choose a security model that split into layers”  (P12 ) VS
“It can be a security risk to include layers of security.” (P1)  \\
Lack   of   standardisation and guidelines  &Lack of standardization & “Standardization  is not really used for now.” (P5) \\
 &Lack of guidelines & “ We do not apply container security guidelines in our company, but
we have a kind of generic guidelines.” (P12)  \\
Unclear resilience time &Undefined resilience time& “ It will depend on the product’s nature and the customer. It could be anything from one or two days to one or two years.” (P7)  \\
 &Average resilience time & S“ We don’t need more than one day on average to resolve the issues.” (P9) \\
Container security is conditional &Conditional security & “I think if all the considerations and risk points are taken containers can be secure.” (P2) \\
 &Unconditional security & “Containers   have  software  delivery  mechanisms,  they  are  secure
enough.” (P7) \\
\end{tabular}
\caption{Summary of Thematic Analysis (Study 1)} 
\label{A summary of the Thematic Analysis Study 1}
\end{table}

\subsubsection{Experience-Based Knowledge}

Experience-based knowledge in this study context refers to practical knowledge gained through exposure to the complexities and challenges in containerized projects. The analysis of interview data highlights a significant variation in practitioners' comprehension of container security. practitioners' knowledge is shaped by individual experiences and the specific demands of their respective fields, not by academic or educational base. 

Interestingly, professional discussions about container security applications were quite different, even within the same domain. Some interviewees assumed that containers were secure by default, and their evidence was that they had not personally encountered security issues. Others assumed it was challenging, and their evidence was many incidents they faced during their work. Most discussions about container security discussed technical aspects, while few prioritize security concerns specific to their domain. Configuration and network vulnerabilities were frequently mentioned as key security risks. 

\subsubsection{Container Security as a Chain of Dependencies}

Practitioners comprehend container security as a life-cycle process, where each phase influences the security of subsequent stages. Consequently, security measures implemented at one stage directly affect the overall security posture of containerized applications.  
Securing containers requires diverse expertise in coding, cloud maintenance, and network security to protect the entire life-cycle. Practitioners emphasize expertise collaboration to clarify and plan interdependent configurations in container deployment. Many vulnerabilities remain undetected until deployment, often revealing issues through irregular system behaviour in production. Therefore, collaboration among experts during the deployment phase plays a crucial role in ensuring a secure and stable production environment.

\subsubsection{Preferring Automation}

Automation in container systems aims to eliminate human involvement in managing, monitoring, and orchestrating containers. According to our findings, security issues often arise from poor configurations in the container development life-cycle. To mitigate risks, practitioners advocate automated solutions on manual configuration to maintain container security. While tasks like base image selection may require manual input, automation can enhance orchestration setup, CI/CD pipeline building, system monitoring, and testing, reducing human errors.

\subsubsection{Common Understanding of the Security Issues}

Container security issues involve risks and vulnerabilities that can arise at any life-cycle phase. Risks can be attacks that exploit system weaknesses, while vulnerabilities arise from design flaws or misconfiguration. We noticed that practitioners have shared knowledge and a deep understanding of the major categories of risks and vulnerabilities in container systems, including, image, host, intra-container, network, and runtime. This knowledge helps reduce the likelihood of risk and potential vulnerability exploits. Additionally, practitioners had similar opinions about most recurring security issues.  They agreed that misconfiguration issues are the most recurring, and they pose a significant threat to container security. 

\subsubsection{Non-Technical Causes}

 Practitioners were aware of technical triggers for container security issues, emphasizing how misconfiguration can lead to security breaches, as anticipated. Surprisingly, some practitioners have pointed out other non-technical factors, such as inadequate team communication and poor organizational and project management.  Practitioners believe that 
 non-technical factors can contribute to security vulnerabilities. They believe that management challenges pose risks as well as technical challenges. While development and deployment issues can be addressed once identified, problems such as team miscommunication or balancing the technology stack with project requirements within the constraints of the customer's budget are more complex.'

\subsubsection{Reliance on Tools}

Practitioners employ many tools across various phases of the container life cycle. Practitioners utilize various open-source, licensed, and proprietary in-house tools. However, practitioners have a heavy reliance on tools, and many of them are not satisfied. 
Some practitioners also presented their companies' future plans to improve security tooling strategies to elevate security levels. Moreover, practitioners emphasized the caution of human administration in tool management. While tools perform their designated function, the results depend on the understanding of the tools' capabilities and their proper implementation and administration.

The tools employed are used for purposes including code quality, identifying vulnerabilities in container images, and managing dynamic and static scanning of container systems. Tools serve to mitigate vulnerabilities during both the building and deployment phases. Additionally, practitioners utilize tools to manage infrastructure configuration and define infrastructure through declarative configuration files. Monitoring tools also track system metrics and behaviour and visualize system data on the front end.

\subsubsection{Uncertainty about Improving Security Practices}

Security-improving practices aim to enhance overall system security without addressing specific issues, unlike mitigation techniques that focus on resolving particular problems. Common practices include selecting secure images, controlling authentication, monitoring network traffic, and managing container development and deployment.

Upon deep analysis of the application of security practices in the container development life cycle, we noticed a conflict in understanding some security practice outcomes in container systems. An example is using container volumes, which are external storage, to save sensitive files. Practitioners supporting this practice claim it is important to store sensitive services away from containers to avoid the implications of unauthorized access. In contrast, practitioners against it claim that using volumes increases the system complexity of the threat tree.

\subsubsection{Lack of Standardization and Guidelines}

Practitioners complained about the lack of documents describing the best practices and systematic protocols for container deployment. They explained their complaint that the available security guidelines are general and do not consider the container infrastructure in terms of sharing resources, and the dynamic nature of containers. Moreover, the security tools and orchestration platforms' best practices and tools are rarely available. Practitioners also emphasized the need for inter-organizational standards for implementation and deployment processes, automating CICD pipelines, disaster recovery plans, and security policies.  

 \subsubsection{Unclear Resilience Time}

Resilience time refers to the duration required to address container security issues. Many practitioners noted that it is difficult to specify a fixed time frame for resolving such issues. It depends on the nature of the security issue and the required experience. However, some practitioners estimated the acceptable time-frame 4 hours to one day. The inability to determine a precise resilience time impacts the security and stability of the system. 

\subsubsection{Container Security is Conditional}

Practitioners deeply believe that containers can be secure enough to support software deployment. At the same time, they put conditions in place to ensure security, like embedding security as an initial element of the development and maintaining good human administration for the container system.

\subsection{Study 2: Key Enablers for Managing Container Security}
\label{Key enablers for container security}

Effective container security management necessitates the identification of key enablers that drive continuous security improvements. Recognizing these enablers provides a foundation for strengthening security measures and ensuring ongoing enhancements in containerized environments. This subsection explores the critical enablers for improving container security, categorizing them into technical and non-technical factors to provide a comprehensive perspective on security advancements in containerized projects. All the themes, codes, and a sample of the quotes are summarized in table \ref{A summary of the Thematic Analysis Study 2}. A consolidated spreadsheet with data analysis, including all the themes, codes, and quotes, is available at: \url{https://zenodo.org/records/14884069}.

\begin{table}[htp]
\centering
\scriptsize
\begin{tabular} {p{1.5cm}p{2cm}p{2.5cm}p{8cm}}

\textbf{Categories} &\textbf{Theme} &\textbf{Example Codes} & \textbf{Example  Quotes }\\
Technical Enablers & Risk Identification & Risk Identification in Container Systems & "Risk identification involves looking for updated packages and base images then refer to public CVEs" (I17)   \\
 &  &Main Challenges in Risk Identification & "Dynamic nature of containers and orchestrator complexity" (I19) \\
  &  &Security Practices support Risk Identification & " We use scanners to detect the risks of the current image that are tied to dependencies, so we can make quick updates" (I14) \\

& Container Testing &Testing types  & "Stress testing is crucial. If you create an API, you should ensure it works as expected and doesn't give out unwanted information." (I4) \\
&  &Challenges in Container Testing  & "Testing itself grows very complex in this dynamic environment, as does creating the test environment setup" (I6) \\
 &  &Testing Can Improve Security Practices  & "Testing results give you confidence that everything works as intended if combined with security practices." (I6) \\

 & logging and Monitoring &  Role of logging and Monitoring &"I think logging and monitoring are crucial. They provide the baseline for issue detection." (I3) \\ 
  &  & Logging and Monitoring Affect Container Security& "The logging and monitoring system is really important to prevent unauthorized access." (I10) \\ 
  &  & Logs Reliability & "Practices like centralized logging, log persistence, policy enforcement, and monitoring contribute to ensuring a certain level of security." (I3) \\ 
   &  & Logging and Monitoring Guide Future Improvements in Container Security &"Seeing issues in the log helps improve practices by making firewall rules stricter and control access." (I1)\\ 
 
  &Artificial Intelligence  & AI helps in Knowledge Sharing & "There could be some kind of tool that gathers discussions and forms a list of requirements of what is needed and what has been discussed. " (I7) \\ 
   &  & AI and Humans  & "Soon, many tasks will be automated many tasks and creative work can be left to humans." (I8) \\ 
    &  & AI Helps in Automation & "AI can enhance container security by automating tasks such as vulnerability scanning, anomaly detection, and threat intelligence." (I16) \\ 
     &  & AI Helps in Testing and Analysis  & "AI could in providing an overall project status, understanding dependencies, and identifying main issues to resolve." (I4)\\ 
      &  & Limitations and Concerns of AI  & " There could be numerous risks that AI poses that our current security measures are not able to address or identify." (I5) \\ 
      
Non-Technical Enablers & Improving the knowledge about Container Security & Improving the knowledge about automation  & "Better knowledge on container security management automation is needed. Utilizing AI and predictive scaling could be future improvements." (I6) \\ 
 &  & Improving the Knowledge about Tools and Best Practices  & "It would be useful to know the characteristics of tools, such as the effort required to integrate them and their benefits." (I1)\\ 
  &  & Improving the Standards and Guidelines & ". Standards for security in containers are very limited. Companies have their own ways, but they should follow high-level standards like ISO." (I12) \\ 
   &  & Common Shared Knowledge about Container Issues & "There are many databases for vulnerabilities, but not everyone uses them." (I1) \\ 

  & Human Collaboration and Communications  & Importance of Human Collaboration & "Good communication is essential. Sometimes when working in a team, we might not see every line of code in a pull request. If we miss important aspects, it could lead to vulnerabilities." (I18) \\   
  
 &  & Maintaining Human Collaboration & " All practices involve human capital, whether it's monitoring, logging, deploying, or fixing attacks. That is why it has to be maintained." (I5) \\ 
 
\end{tabular}
\caption{Summary of Thematic Analysis (Study 2)} 
\label{A summary of the Thematic Analysis Study 2}
\end{table}

\subsubsection{Technical Key-Enablers}

Technical enablers are the technology-related factors that support and enhance container security in containerized projects. The analysis identified five technical enablers: risk identification, testing, logging and monitoring, security practices, and AI.

\paragraph{\textit{Risk identification}}

Risk identification refers to recognizing and addressing potential vulnerabilities and threats that affect container systems. Risk identification can be achieved by detecting abnormal system behaviour through tools or by combining both. Effective risk identification in software containers involves continuous updates, anomaly detection strategies, and tooling plans.

Risk identification in container systems faces many challenges. These challenges arise from the complexity of container systems design and the combination of static and dynamic elements operating together. One of the main challenges in risk identification is the continuous need for updated security tools and the need for effective security tool administration to address both known and unknown anomalies. Container system complexity is also another challenge in container systems. The integration between containers, Hosting machines, orchestration platforms, and user inputs makes it hard to trace the security issues in the threat tree.

\paragraph{\textit{Container Testing }}
Container testing is essential for ensuring containers' security and performance. Although testing increases the workload on development teams, it is crucial to identify and mitigate security risks before they can affect the entire system.  Container systems require different types of testing to ensure a secure performance for container systems, such as unit testing, integration testing, stress testing, and end-to-end testing. Unit testing involves testing that individual components are functioning properly on their own. Integration testing focuses on verifying the integration of container components. Stress testing checks the performance stability under a heavy workload. End-to-end testing ensures that the application will perform as expected in a production environment.

Container testing faces many challenges in container systems. One of the challenges that most industries are facing is that developers are taking responsibility for container applications they are developing instead of a separate testing team due to budget constraints. This puts extra load on the developers, as they must test the functionality in addition to security without a clear knowledge of the security metrics. Another challenge is the difficulty of managing testing within a large number of containers at the same host, making it hard to test and audit the container dependencies effectively.

Testing results of containerized applications offer valuable insights for enhancing security. 
Testing results help to provide statistics on recurring security threats in container systems. The identified threats from the testing process should be thoroughly examined to determine effective mitigation and recovery strategies. These results from testing processes should be considered and translated into practical security improvements in container systems.

\paragraph{\textit{Security Practices} }

Security practices in container systems are a comprehensive set of planned actions and strategies that aim to protect applications, infrastructure, and data in a container environment. Security practices are supposed to be proactive and continuous to mitigate the threats in their early phases.  Proactive security practices must cover various levels, including image, code, application, infrastructure,  ports, nodes, and user inputs. 

 In addition to proactive security practices, strategic security practices are essential for ensuring the protection and integrity of container systems. It must include an integrated set of procedures, tools, and strategies to protect the container system. Strategic security practices ensure that security is embedded into the development process and aligns with DevSecOps principles for maximum protection. Strategic security practices should be tailored to the specific needs of each project to ensure security is effectively integrated into the development life-cycle.

Following proactive and strategic security practices provides a structured approach to managing security in container systems. It provides a plan for implementing security in container systems, that can be tailored according to the customer's needs and available budget. Moreover, it schedules defined time for regular auditing and vulnerability assessments that minimize threat exposure. 

\paragraph{\textit{Logging and Monitoring}}
Logging and monitoring are continuous processes of collecting and analyzing data about the system's behaviour, including errors, activities, resource consumption and performance. Logging and monitoring support container security in many ways. It helps to identify security issues and track their origin source, whether it is a user, system element,  or container application. It provides a real-time overview of the running system to evaluate threats before it extends to other system elements. Moreover, it alerts the security team about users' failed logging attempts and the credentials that are exposed in log files.

Unauthorized access is one of the major risks that affect the reliability, integrity and confidentiality of the container logs. Therefore, Logging and monitoring security practices such as access control and encryption are essential to mitigate unauthorized access. Ensuring the reliability of the logs' data requires continuous updates to the container systems dependencies, security practices, and security tools.

\paragraph{\textit{Artificial Intelligence } }

Artificial intelligence (AI) plays an important role in improving container security. It can be embedded in various security aspects of container systems. One of the main aspects of AI being a key player in container security is testing. Tools like Docker Scout \footnote{https://docs.docker.com/scout/} and Trivy \footnote{https://trivy.dev/latest/} are using AI for unit testing in addition to their original function as vulnerability scanners. AI helps runtime security by detecting abnormal behaviour in container logs. 

AI can strongly support security management in container systems. AI can automate security practices to avoid human mistakes, for example, AI can automate YAML files---human-readable data serialization format used for configuration files--- and image code modification according to the security guidelines.  AI can also be used to monitor, assess risk, and check logs to detect patterns or anomalies. Some AI tools like CrowdStrike \footnote{https://www.crowdstrike.com/platform/cloud-security/} help support anomaly decisions by prioritizing vulnerabilities in container systems. AI is also used to enforce security policies; for example, Aqua Security \footnote{https://www.aquasec.com/} is used to enforce container runtime security policies and block unauthorized processes in container systems.

AI helps improve communication and knowledge sharing among the development team. It can improve project collaboration and communication by summarizing meetings, tracking progress, transcribing discussions and creating project documentation. AI can also help at the foundational level for new employees' onboarding by mimicking a trial-and-error environment and giving guidance when needed to enhance learning and implementation. 

Despite the significant benefits AI can introduce to container security, it is associated with serious concerns. One of the main concerns is that developers use AI heavily in code generation, which introduces the potential to generate malicious codes or expose sensitive data. Another concern is that the effectiveness of AI security solutions depends on the expertise of their users. If the user lacks experience, the AI security solutions outputs may be misinterpreted or improperly applied, leading to security vulnerabilities.
Hence, AI should be used as a tool under the supervision of experienced professionals to ensure the reliability of security solutions.

\subsubsection{Non-technical Key-Enablers}
The non-technical enablers are the human-based enablers that help improve container security in containerized projects. The data analysis provided two main non-technical enablers for container security: knowledge sharing and human collaboration and communication. 

\paragraph{ \textit{Sharing knowledge about Container Security}}

Enhancing knowledge sharing in container security requires a deeper understanding of the challenges that practitioners face in container security. One of the challenges that require more knowledge sharing in container systems is tools and their best practices. Improving knowledge about tools and their best practices needs trusted and comprehensive resources. While courses are usually recommended to learn more about tools and their best practices, they can help with foundational knowledge. For more advanced knowledge, practitioners recommend reading project documentation that provides valuable knowledge on tools and their best practices. 

Another concern that requires increased knowledge sharing is the secure implementation of automation in container systems. Practitioners recommended a balanced approach to applying secure automation in container systems, where routine tasks can be automated combined with human administration and monitoring. Furthermore, practitioners emphasize the importance of sharing knowledge to automate tasks such as downloading libraries, managing caches, and setting up initial infrastructure to reduce manual verification.

An alternative approach to enhancing shared knowledge about vulnerabilities in container systems is utilizing open-source vulnerability databases. Vulnerability databases such as CVE \footnote{https://www.cvedetails.com/} and Snyk \footnote{https://security.snyk.io/}help to identify, track, and mitigate vulnerabilities within container systems.  In addition to vulnerability databases, there are other sources of knowledge about vulnerabilities, such as workshops, webinars, committee meetups, and recorded videos. 

Another effective way to share knowledge about container security is through training programs, where professionals share personal experiences, lessons learned, and insights into how these experiences have influenced their approaches to implementing security in containers. Practitioners believe that these valuable insights should not be confined to training programs only. Instead, they should be shared more broadly through collaborative platforms, blogs, and documented use cases to make knowledge accessible to a wider audience.

\paragraph{\textit{Human Collaboration and Communication} }

Effective human communication and collaboration are essential for successfully implementing and managing container security. Collaboration among teams, including developers, network engineers, cloud experts and hardware specialists, ensures that all elements of the container system work cohesively. Clear communication regarding the implementation and configuration of each phase in the container life-cycle helps teams avoid potential configuration inconsistencies and mitigate potential security risks.

Maintaining human communication and collaboration requires regular team meetings as well as accessible communication channels. Regular meetings, whether daily scrums or weekly sync-ups, help keep everyone informed about ongoing tasks and issues. Accessible communication channels are also essential for individual discussions. Industries use various communication channels, such as Slack \footnote{https://slack.com}, Microsoft Teams \footnote{https://www.microsoft.com/en-us/microsoft-teams}, and WhatsApp groups \footnote{https://www.whatsapp.com}, to ensure that all teams are updated on urgent matters. Human collaboration should be actively encouraged to ensure that all team members are involved and contribute towards a secure container system implementation.

\section{Discussion}
\label{Discussion}

This research explores security management in containerized projects, examining two primary aspects: practitioners' perceptions of container security and the key enablers for enhancing security in containerized projects. The findings emphasize technical and non-technical patterns in addition to technical and non-technical enablers. Integrating technical patterns and enablers with non-technical factors provides a holistic approach to managing container security. This integration ensures a more comprehensive and sustainable security strategy, fostering proactive threat mitigation and continuous improvement in security practices.



A thorough analysis of our findings reveals that practitioners and researchers largely agree on the expected challenges and strategies for improving container security. Both groups recognize and confirm the importance of technical enablers such as anomaly detection, AI, security practices, testing, logging, and monitoring—in managing container security. Furthermore, there is agreement on the importance of non-technical enablers, such as improving container security knowledge and fostering human collaboration within containerized projects. However, differences emerge regarding the implementation of these improvements.  For example, practitioners' expectations about the role of AI in securing containers are limited to automation and replacing human tasks in analysis and testing, while researchers' ideas were more focused on using AI in knowledge sharing, supporting human collaboration, and risk management. Another example of AI concerns and limitations is that practitioners think there is no harm in using AI coding. In contrast, researchers think using AI in coding might introduce hidden risks in the codes that can cause data leakage.

The data analysis further reveals that improving container security requires increased collaboration from leading industries. These industries must maintain transparency by sharing internal standards and guidelines for managing containerized projects. There is also a need for more collaboration across industries to standardize processes for employee training and develop security guidelines suitable for general applications. Additionally, industries must engage more actively with regulatory authorities to ensure that future container security advancements align with legal data protection requirements.

Although improving the knowledge about container security is a personal responsibility for the practitioners in the first place, we think that it is also the responsibility of industries interested in containers. Individuals should proactively seek to attend relevant courses and webinars and read research articles and blogs on container security. industries also have responsibilities towards improving employees' knowledge about container security by providing fundamental and advanced training, proper onboarding, and supporting resources and tools. Moreover, organizations interested in containers establish alliances between industries and research institutes to deliver improved processes, standards, workshops, platforms, and YouTube channels to provide free knowledge about container security.

\begin{figure}
    \centering
    \includegraphics[width=1\linewidth]{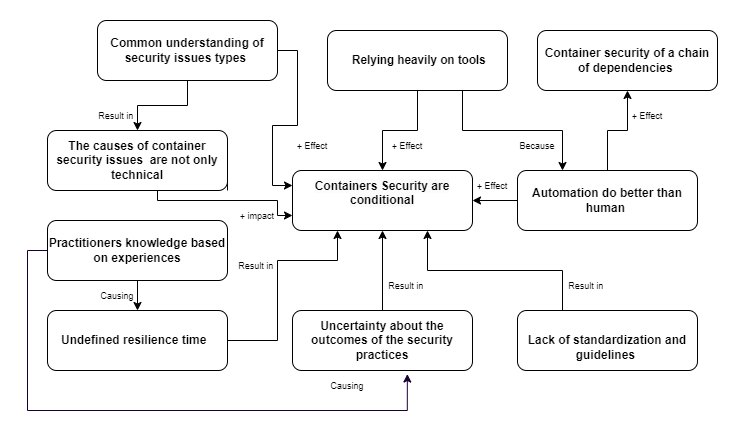}
    \caption{Model 1: Container security pattern interrelation model}
    \label{fig:model}
\end{figure}

Fig. \ref{fig:model} illustrates our initial model that describes the interconnections among the themes on how practitioners perceive the issues and challenges in container security. Container security faces several challenges, including uncertainties in enhancing security practices, time limitations for resolving vulnerabilities, and the lack of standardized guidelines. Automating security processes and implementing security tools are crucial in strengthening container security by reducing human errors. Additionally, establishing a shared understanding of security issues is essential, as it provides deeper insights into their root causes and impact on container systems, thereby facilitating effective risk management. Practitioners also acknowledge the role of non-technical factors, such as efficient project management, in aligning security measures with the existing technology stack. Moreover, the model also emphasizes the influence of project-based experience on developing security expertise, reflected in variations in the time required to address security issues and differing perspectives on enhancing security practices.


By establishing the relationships between these themes, we can focus on the strengths and weaknesses of container security practices. Strengths include a comprehensive understanding of security issues, reliance on tools and automation, awareness of security dependencies, and consideration of non-technical factors. Conversely, weaknesses encompass the lack of systematic knowledge, guidelines, and standards, uncertainties regarding practice improvements, and resilience time. Integrating the enablers for improving security with the security patterns from the first study will provide a more comprehensive understanding of how security can be managed in containerized projects.

Thus, we developed our revised model (as shown in Fig. \ref{Key Improvements in Container Security}), which provides an overview of how practitioners manage security in containerized projects. The model further addresses the essential aspects and improvements and the relevant patterns as follows:

\begin{figure}
    \centering
    \includegraphics[width=1\linewidth]{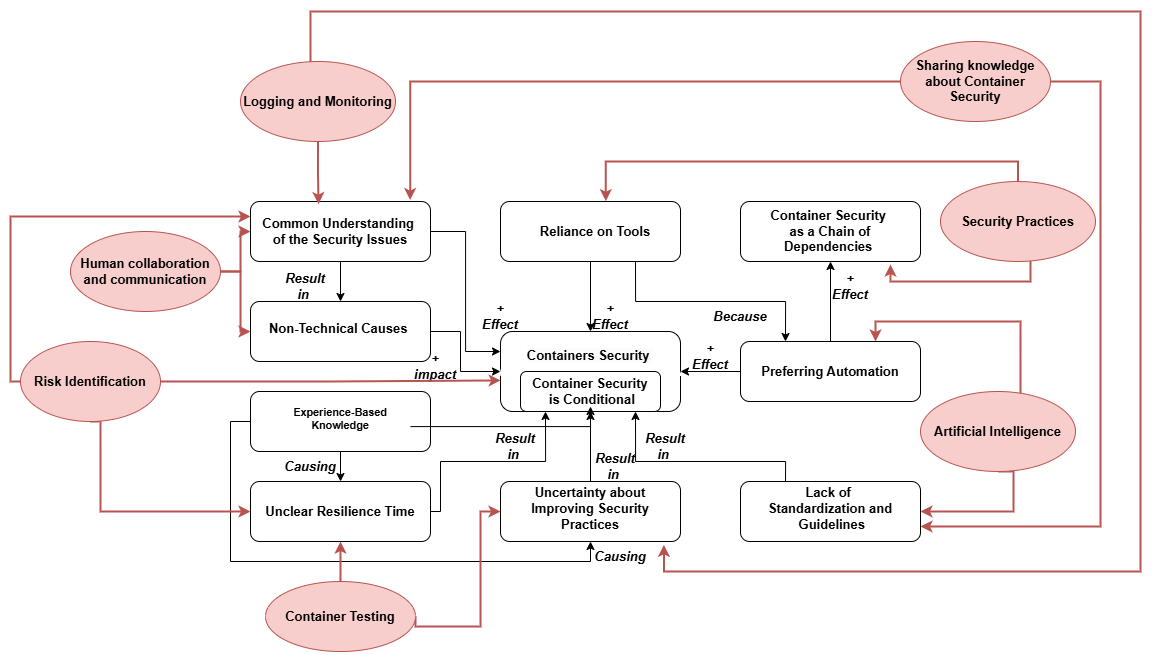}
    \caption{Model 2: Managing Security in Containerized Project}
    \label{Key Improvements in Container Security}
\end{figure}

\textit{Model 2} visualize how security is managed in containerized projects. The red arrows denote the specific key enablers addressing the relevant security patterns. Below, we describe in detail the process.

\textit{Risk Identification} supports container security by identifying potential risks in container systems during the early phases of deployment. Analyzing risk identification techniques and their impact on the system will help estimate the time required to resolve security issues.

\textit{Container Testing} reduces uncertainty in security practices by providing empirical evidence of their effectiveness when a system successfully passes various security tests, such as unit testing, integration testing, stress testing, and end-to-end testing. Additionally, testing helps determine resilience time by accurately evaluating the time required to detect, analyze, and mitigate vulnerabilities in the testing process.

\textit{Security Practices} supports the container supply chain in container systems by safeguarding the chain of dependencies throughout container development, and deployment phases. Moreover, applying security practices can significantly improve the performance of security tools, as it will substantially reduce false positive alerts in security reports.

\textit{Logging and Monitoring} provides a detailed record of security events. Sharing these records among team members fosters a common understanding of security issues and the necessary precautions. Furthermore, continuous logging and monitoring help mitigate uncertainties regarding security practices by offering real-time insights into potential vulnerabilities.

Integrating \textit{Artificial Intelligence} into container system development and deployment enhances security by automating vulnerability scanning and ensuring configuration compliance. AI can also compensate for the lack of general standardization for the development processes by analyzing the effect of security practices on overall container security and prioritizing effective practices.

\textit{Sharing Knowledge about Container Security} encourages the exchange of best practices and experiences, fostering a shared understanding of security issues among developers about container security. Additionally, sharing internal security standards among industries and organizations interested in container security will contribute to a collective repository of data that can inform future guidelines for securing container systems.

the exchange of best practices and experiences, fostering a shared understanding of security issues among all developers.

\textit{Human Collaboration and Communication} is one of the primary non-technical factors influencing container security. Effective communication within teams enhances their understanding of security challenges and promotes collaborative efforts to address them during the development phase.

\subsection{Implications on SE Practice}

 Our findings contribute to SE practices by improving and managing container security issues in the following ways:

\begin{enumerate}
\item Industries should focus on structured guidelines for project documentation to ensure that all relevant security measures, configurations, and challenges are recorded. Documentation should be regularly updated to reflect the current status of the project and any emerging security concerns. Maintaining structured and updated documentation facilitates knowledge retention within teams and streamlines the onboarding process for new developers.
\item  Industries can highly benefit from AI security solutions in containerized projects such as automated threat detection, and self-healing systems. AI security solutions can significantly enhance system resilience, and detect security breaches before they escalate. Moreover, it enables the security teams to focus on more complex threats and strategic security planning.

\item Industries need to continue investing in advanced security tools and ensure tools are visible to the team to maintain security in containerised projects. Advanced security tools need to be implemented alongside practical security training.  Security training must be tailored to developers for effective tool utilization. Industries need to incorporate workshops, interactive simulations, and real-world attack scenarios to help employees develop security skills. 

\item  The incorporation of gamification in security training can enhance engagement and knowledge retention, particularly for early-career developers. Security training programs should include interactive techniques such as real-world threat simulations with multilevel security challenges to develop a proactive security mindset. These techniques will help developers master security practices and apply them to real projects.

\end{enumerate}

\subsection{Future SE Research Avenues}
Building upon the findings of this study, several avenues for future research can further advance the understanding of container security. The following research directions are proposed:

\begin{enumerate}  
\item  Establishing standardized security metrics is essential to evaluate security in containerized environments. Future SE research in containers should focus on defining measurable indicators that facilitate effective risk assessment, resource allocation, and mitigation strategies. An empirical approach can provide insights into the most critical security concerns that require prioritization.

\item  Future studies should also assess the implications of security measures in different containerized environments. Employing exploratory research will help identify domain-specific security priorities and best practices.

\item  The responsibility of DevOps teams in securing container systems needs more exploration in the container context. Investigating ownership and accountability of security issues in security management can provide insights into how policies influence security outcomes. A mixed-methods approach ---combining qualitative research and quantitative --- can offer a comprehensive understanding and validate ownership and accountability in containers.

\item  Future research should examine the ethical implications of AI security tools in container systems, particularly regarding the exposure of personal and sensitive data. Multidisciplinary research, including legal analysis, ethical frameworks, and technical evaluations, can provide a balanced perspective on the ethical implementation of AI in container security.

\end{enumerate}

\subsection{Threats to Validity}
\label{Assessing Quality Criteria}
To ensure the rigor and trustworthiness of our study, we refer to the ACM SIGSOFT Empirical Standards \cite{ralph2020acm} in addressing the quality criteria of our research. 
\begin{itemize}

\item \textit{Credibility}: We maintain the credibility of the results by including supporting quotes for each identified theme. It also supports the reproducibility of the themes.  A consolidated document with all the direct quotes, codes, and themes is available at: \url{https://zenodo.org/records/10959273} and \url{https://zenodo.org/records/14884069};

\item \textit{Usefulness}: the findings of this research benefit practitioners by offering a visual model of container security management. The model highlights the patterns in containerized projects and the enablers to improve these patterns. 

\item \textit{Transferability}: the model describing the security patterns and their relationship summarizes the experiences of practitioners working across various domains and roles. Additionally, it connects each pattern to the specific improving enabler. This makes the results comprehensive and applicable to a wide range of projects and domains;

\item \textit{Resonance}: we explain the strengths and weaknesses in the security patterns of container systems and provide an enabler to deepen the understanding of security management in container systems. Software practitioners could directly use these data to enhance, maintain, and manage container security. 

\end{itemize}

\section{Conclusion}
\label{Conclusion}

Software containers have become a widely adopted approach for efficiently deploying software-intensive applications. However, existing SE research literature on security management predominantly focuses on technical security practices and testing methodologies, neglecting the significant role of human administration in planning, decision-making, and strategy development container systems. Consequently, this research contributes to the knowledge of security management in container systems by highlighting how SE practitioners perceive the various security challenges and their approaches to managing these security issues in software container systems.  


We conducted two semi-structured interview studies to examine how practitioners manage security issues. While the first study explored how practitioners perceive security issues in containerized systems regarding their causes and implications, the second study investigated how SE practitioners manage these security issues in containerized projects. 

The following are the main findings from our research: 

 \begin{enumerate}
     \item Our findings provide insights into how practitioners perceive security issues, their causes, and the mitigation techniques and provide an overview of the security patterns in containerized projects.     
     \item The findings also present the advances of containers as a solution for deploying software applications in terms of clarity of security issues, integrating tools that help improve security and automation, and consideration of non-technical factors during developing and deploying containerized systems. 
     
  \item The findings also explore the weaknesses of containerized software systems, including the lack of systematic knowledge about security issues, guidelines uncertainty regarding practice improvements, and undefined resilience time.
  
     \item Furthermore, we identified key enablers for improving container security, categorizing them into technical and non-technical factors. Technical enablers include risk identification, security testing, security practices, logging and monitoring, and AI solutions. Non-technical enablers encompass knowledge sharing, effective communication, and collaboration among team members. A combination of technical and non-technical enablers ensures comprehensive improvements in container security on the technical and strategic levels.  
     
     \item We propose a conceptual model that describes how practitioners manage security in containerized projects. The model presents the security patterns, illustrates their interconnections, and highlights key enablers that support effective security management. The model will guide practitioners in developing robust strategies for planning and deploying highly secure container systems.
 \end{enumerate}

\section*{Acknowledgements}
\label{Acknowledgements}
This research is supported by \textit{Containers as the Quantum Leap in Software Development (QLeap)} project funded by Business Finland (BF) grant; number 3215/31/2022. 

During the preparation of this work the author(s) used Copilot in order to enhance the readability and clarity of the text. After using Copilot, the author(s) reviewed and edited the content as needed and take(s) full responsibility for the content of the publication.

\bibliographystyle{elsarticle-num-names} 
\bibliography{citation}

\end{document}